\documentclass[12pt,a4paper]{article}
\usepackage{amsmath,caption,amsfonts,setspace,mathrsfs,mathtools}
\usepackage[top=1.2in, bottom=1.2in, left=1in, right=1in]{geometry}

\usepackage[round]{natbib}
\usepackage{color,soul}

\DeclareCaptionStyle{italic}[justification=centering]{labelfont={bf},textfont={it},labelsep=colon}
\usepackage{graphicx,psfrag,epsf,textcomp}
\usepackage{enumerate, dsfont}
\usepackage{natbib}
\usepackage{url,xcolor} 
\usepackage{booktabs, subfig, bm, paralist,mathpazo,tikz,todonotes,longtable,microtype}

\usepackage[pdftex,colorlinks=true]{hyperref}
\definecolor{darkblue}{rgb}{0,0,.6}
\hypersetup{citecolor=darkblue,linkcolor=darkblue,urlcolor=darkblue}

\date{}

\newcommand{\blind}{0}

\graphicspath{{plots/}}

\newsavebox\CBox

\begin{document}

\def\spacingset#1{\renewcommand{\baselinestretch}{#1}\small\normalsize} \spacingset{1}

\if0\blind
{
  \title{\bf Bayesian bandwidth estimation for local linear fitting in nonparametric regression models}
  \author{
    Han Lin Shang\footnote{Postal address: Department of Actuarial Studies and Business Analytics, Level 7, 4 Eastern Road, Macquarie University, Sydney, NSW 2109, Australia; Telephone: +61(2) 9850 4689; Email: hanlin.shang@mq.edu.au; ORCID: \url{https://orcid.org/0000-0003-1769-6430}}\\
Department of Actuarial Studies and Business Analytics \\
Macquarie University \\
\\
    Research School of Finance, Actuarial Studies and Statistics \\
 Australian National University  \\
 \\
 Xibin Zhang \\
 Department of Econometrics and Business Statistics \\
 Monash University
 }
  \maketitle
} \fi

\if1\blind
{
    \title{\bf Bayesian bandwidth estimation for local linear fitting in nonparametric regression models}
  \maketitle
} \fi

\begin{abstract}
This paper presents a Bayesian sampling approach to bandwidth estimation for the local linear estimator of the regression function in a nonparametric regression model. In the Bayesian sampling approach, the error density is approximated by a location-mixture density of Gaussian densities with means the individual errors and variance a constant parameter. This mixture density has the form of a kernel density estimator of errors and is referred to as the kernel-form error density \citep[c.f.,][]{ZKS11}. While \cite{ZKS11} use the local constant (also known as the Nadaraya-Watson) estimator to estimate the regression function, we extend this to the local linear estimator, which produces more accurate estimation. The proposed investigation is motivated by the lack of data-driven methods for simultaneously choosing bandwidths in the local linear estimator of the regression function and kernel-form error density. Treating bandwidths as parameters, we derive an approximate (pseudo) likelihood and a posterior. A simulation study shows that the proposed bandwidth estimation outperforms the rule-of-thumb and cross-validation methods under the criterion of integrated squared errors. The proposed bandwidth estimation method is validated through a nonparametric regression model involving firm ownership concentration, and a model involving state-price density estimation. 

\vspace{0.2cm}

\noindent Keywords: kernel-form error density, Markov chain Monte Carlo, ownership concentration, state-price density


\end{abstract}

\def\spacingset#1{\renewcommand{\baselinestretch}{#1}\small\normalsize} \spacingset{1}
\spacingset{1.52}

\newpage

\section{Introduction}

Local linear fitting is an important and widely used nonparametric tool for exploring an unknown relationship between a response variable and a group of explanatory variables. The local linear estimator reduces estimation error compared to the local constant estimator. While the choice of kernel function is not important, the performance of the local linear estimator is mainly determined by choice of bandwidth \citep[see, e.g.,][p.76]{FG96}. Similar to a previous study by \cite{ZKS11}, we present a Bayesian sampling approach to bandwidth estimation for a nonparametric regression model with unknown error density. However, in contrast to \cite{ZKS11, ZKS16}, we consider local linear fitting rather than local constant fitting, to improve the estimation accuracy of the regression function.

A large body of literature discusses methods for bandwidth selection, such as the rule-of-thumb (ROT) and cross-validation (CV) discussed by \cite{Hardle90} and \cite{Scott92}, the plug-in method discussed by \cite{HEW+95}, and bootstrapping as proposed by \cite{HLP95}. In terms of the local constant estimator, a particular version of the local linear estimator, \cite{ZKS11} present a simulation study, finding that CV outperforms ROT and bootstrapping. As evidenced by the simulation results presented in this paper, CV also outperforms ROT for local linear fitting. This finding partly explains why CV is often used in a wide range of application studies. However, even if CV is asymptotically unbiased, it has a relatively large finite-sample bias. \cite{FJ90}, \cite{PT86} and \cite{Romano88} have established the same behavior studying nonparametric estimates for the density, quantiles and the mode, respectively. This problem arises in nonparametric-based models, as exemplified by \cite{HM93}. In addition, CV sometimes fails to produce meaningful results. \cite{LZ05} show that in terms of the local constant estimator with one regressor, a necessary and sufficient condition has to be met by the CV to guarantee the existence of a unique optimal bandwidth. The conditions required to ensure the uniqueness of a cross-validated bandwidth vector become complicated where they are multiple regressors and/or local linear fitting. Moreover, in application studies, it is difficult to verify such conditions. Therefore, the lack of an applicable data-driven bandwidth selection method for local linear fitting motivates the investigation carried out in this paper.

Existing bandwidth selection methods do not take error density into account. Although these methods do not require any information about the error distribution, the distribution of the response variable is often of great interest to error-based techniques, such as value-at-risk estimation \citep[e.g.,][]{MYT18}. Such a distribution is characterized by the error density, estimation of which is fundamental in statistical inference for any regression model. A straightforward approach to error density estimation is to use residuals as proxies for errors and derive a kernel estimator. However, bandwidth selection is yet to be investigated when the regression function is estimated through local linear fitting. Thus, the proposed investigation is also motivated by the lack of a data-driven procedure for simultaneously choosing bandwidths in the local linear estimator and the kernel estimator of error density.

Let $\bm{y}$ denote the response and $\bm{x}=\left(x_1,x_2,\dots,x_d\right)^{\top}$ a set of explanatory variables. Given observations $\left(y_i,\bm{x}_i\right)$, the multivariate nonparametric regression model is expressed as
\begin{equation*}
y_i=m(\bm{x}_i)+\varepsilon_i,\qquad i=1,2,\dots,n, \label{regression}
\end{equation*}
where $(\varepsilon_1,\varepsilon_2,\dots,\varepsilon_n)$ are assumed to be independent and identically distributed (iid) with an unknown error density, denoted as $f(\varepsilon)$. It is also assumed that the errors are independent of the explanatory variables.

The regression function $m(\bm{x})$ can be estimated by the local linear estimator denoted by $\widehat{m}(\bm{x},\bm{h})$, where $\bm{h}=(h_1,h_2,\dots,h_d)^{\top}$ is a vector of bandwidths corresponding to the $d$ explanatory variables. We propose to approximate the unknown error density by a mixture density given by
\begin{equation}\label{errordensity}
f(\varepsilon;b)=\frac{1}{n}\sum^n_{i=1}\frac{1}{b}\phi\left(\frac{\varepsilon-\varepsilon_i}{b}\right),
\end{equation}
where $\phi(\cdot)$ is the probability density of the standard Gaussian distribution. It is a location-mixture density with Gaussian components having a common variance $b^2$ and individual means at the errors. The validity of this mixture density as an error density has been investigated in \cite{ZKS11} for the local constant estimator in nonparametric regression models. As $f(\varepsilon;b)$ has the form of a kernel density estimator of errors, it can approximate the true error density very well when the sample size is sufficiently large.

There is a growing literature on the estimation of the error density in a nonparametric regression model. \cite{Efromovich05} presented the so-called Efromovich-Pinsker estimator of the error density and show that this estimator is asymptotically as accurate as an oracle that knows the underlying errors. \cite{Cheng04} demonstrates that the kernel density estimator of residuals is uniformly, weakly, and strongly consistent. While the regression function is estimated by the Nadaraya-Watson estimator, the error density is estimated by the kernel estimator of residuals. \cite{Samb10, Samb11} proved the asymptotic normality of the bandwidths in both estimators, deriving the optimal convergence rates of the two bandwidths. \cite{LX07} propose a kernel estimator based on residuals obtained through the local polynomial fitting of the unknown regression function. They showed that the estimator is adaptive and conclude that adaptive estimation is possible in the local polynomial fitting. In a class of nonlinear regression models, \cite{YD07} constructed an approximate (pseudo) likelihood through the kernel density estimator of pre-fitted residuals with its bandwidth pre-chosen by the rule-of-thumb. They prove that under some regularity conditions, the resulting maximum likelihood estimates of parameters are consistent, asymptotically normal, and efficient. \cite{JW08, JW11} propose using the kernel density estimator of the pre-fitted residuals to construct an approximate likelihood, which they call ``kernel likelihood". In all these papers, residuals are commonly used as proxies of errors, and the bandwidth for the kernel density estimator of residuals is pre-chosen or chosen in a way that is unrelated to the bandwidth in the regression function. To our knowledge, no method can simultaneously estimate the bandwidths for the local linear estimator of the regression function and kernel-form error density.

The main contribution of this paper is to construct an approximate likelihood and, therefore, the posterior of the bandwidth parameters for the nonparametric regression model with the kernel-form error density given by~\eqref{errordensity}, where the unknown regression function is estimated through the local linear fitting. In contrast to \cite{ZKS11, ZKS16}, the local linear fitting improves the estimation accuracy of the regression function. A Bayesian sampling algorithm is presented to sample the bandwidth parameters from their posterior; thus, the two types of bandwidths can be estimated simultaneously. We conduct a Monte Carlo simulation study, which shows that our proposed method outperforms the ROT and CV methods where there are multiple regressors under the criterion of integrated squared errors (ISE). The proposed bandwidth estimation method is also validated through a nonparametric regression model involving firm ownership concentration, and a model involving state-price density (SPD) estimation.

The remainder of this paper is organized as follows. In Section~\ref{sec:2}, we present the likelihood and posterior of bandwidth parameters based on local linear fitting. Section~\ref{sec:3} presents  Monte Carlo simulation studies to examine the accuracy of the proposed bandwidth estimation method. The proposed Bayesian method is validated through a nonparametric regression model involving firm ownership concentration in Section~\ref{sec:4}. In Section~\ref{sec:5}, we apply our method to bandwidth estimation in a nonparametric regression model involved in SPD estimation. Section~\ref{sec:6} concludes the paper.

\section{Bayesian bandwidth estimation}\label{sec:2}

The bandwidths in the local linear estimator of the regression function and kernel-form error density estimator play an important role in controlling the smoothness of the regression function and error density estimator, respectively. In the case of the local constant estimator, bandwidths are treated as parameters by \cite{ZBK09, ZKS11}.

In this paper, the error density of the nonparametric regression model given by~\eqref{regression} is approximated by the kernel-form density $\phi(\cdot)$ in~\eqref{errordensity}. When the error density of a regression model is non-Gaussian, a scale-mixture density of several Gaussian components is often used, where each Gaussian density may have mean zero and different variance parameters. However, the use of a scale-mixture density of Gaussian components is at the cost of dramatically increasing the number of parameters. In contrast, the kernel-form error density has the advantage of having only one parameter, the common variance parameter, which also acts as the bandwidth for this kernel-form density.

\subsection{Cross-validation}

Consider the nonparametric regression model given by
\begin{equation}
y_j=m(\bm{x}_j)+\varepsilon_j, \qquad j=1,2,\dots,n, \label{eq:100}
\end{equation}
where $\bm{x}_j=(x_{j,1},\dots,x_{j,d})$ is a random vector of dimension $d$. Define the derivative of $m(\bm{x})$ as $\beta(\bm{x})=\partial m(\bm{x})/\partial \bm{x}$.

Let $\delta(\bm{x})=\left(m(\bm{x}), \beta(\bm{x})^{\top}\right)^{\top}$ be a vector-valued function with dimension $d+1$. Taking a Taylor series expansion of $m(\bm{x}_j)$ at $\bm{x}_i$, we obtain
\begin{equation}
  m(\bm{x}_j)=m(\bm{x}_i)+(\bm{x}_j-\bm{x}_i)^{\top}\beta(\bm{x}_i)+o_{ij},\label{eq:2}
\end{equation}
where $o_{ij}$ represents the remaining term of the expansion. Substituting~\eqref{eq:2} into~\eqref{eq:100}, we obtain
\begin{equation*}
  y_j=\left(1,(\bm{x}_j-\bm{x}_i)^{\top}\right)\delta(\bm{x}_i)+o_{ij}+\varepsilon_j.
\end{equation*}

As pointed out by \cite{LR04}, the leave-one-out local linear estimator of $\delta(\bm{x}_i)=\left(m(\bm{x}_i),\beta(\bm{x}_i)^{\top}\right)^{\top}$ is obtained by a kernel weighted regression of $y_j$ on $\left(1, (\bm{x}_j-\bm{x}_i)^{\top}\right)$ given by
\begin{align*}
\widehat{\delta}_{-j}\left(\bm{x}_j;\bm{h}\right)=\left[\sum^n_{\substack{i=1 \\ i\neq j}}K_{\bm{h}}(\bm{x}_j;\bm{x}_i)
\left(\begin{array}{cc}
     1 & (\bm{x}_j-\bm{x}_i)^{\top} \\
     \bm{x}_j-\bm{x}_i & (\bm{x}_j-\bm{x}_i)(\bm{x}_j-\bm{x}_i)^{\top} \\
    \end{array} \right)
  \right]^{-1}\sum^n_{\substack{i=1 \\ i\neq j}}K_{\bm{h}}(\bm{x}_j;\bm{x}_i)
  \left(\begin{array}{c}
    1 \\
    \bm{x}_j-\bm{x}_i
  \end{array}\right)y_i,
\end{align*}
for $j=1,2,\dots,n$, where $K_{\bm{h}}(\bm{x}_j;\bm{x}_i)=\prod^d_{k=1}h_k^{-1}K\left((x_{j,k}-x_{i,k})/h_k\right)$ with $h_k$ being the bandwidth corresponding to the $k$\textsuperscript{th} component of $\bm{x}$. Note that $x_{j,k}$ is the $k$\textsuperscript{th} element of $\bm{x}_j$, for $k=1,2,\dots,d$.

The leave-one-out estimator of $\widehat{m}(\bm{x}_j)$ is given by
\begin{equation*}
   \widehat{m}_{-j}(\bm{x}_j;\bm{h})=\bm{I}_1^{\top}\delta_{-j}\left(\bm{x}_j;\bm{h}\right),
\end{equation*}
where $\bm{I}_1$ is a $(d+1)\times 1$ vector, the first element of which is one and other elements zero. An optimal $\bm{h}$ can be chosen by minimizing the CV function given by
\begin{equation}
\text{CV}(\bm{h}) = \sum^n_{j=1}\left[y_j-\widehat{m}_{-j}(\bm{x}_j;\bm{h})\right]^2.\label{eq:50}
\end{equation}
The CV method is considered the benchmark method for comparison with our proposed method (to follow).

\subsection{Likelihood}

\cite{ZKS11} propose approximation of the unknown error density by a location-mixture density of $n$ Gaussian densities, where the component Gaussian densities have a common variance and different means at individual errors. The errors were approximated by the residuals from a nonparametric regression model with the regression function estimated by the local constant estimator. In this paper, we follow the same idea for the nonparametric model with a local linear estimator of the regression function, where the error density is assumed to be approximated by~\eqref{errordensity}, and, therefore, the density of $y_i$ is approximated by $f\big(\{y_i-\widehat{m}_{-i}(\bm{x}_i;\bm{h})\};b\big)$ expressed as
\begin{equation}
\widehat{f}\big(\{y_i-\widehat{m}_{-i}(\bm{x}_i;\bm{h})\};b\big)=\frac{1}{n}\sum^n_{\substack{j=1 \\ j\neq i}}\frac{1}{b}\phi\left(\frac{\{y_i-\widehat{m}_{-i}(\bm{x}_i;\bm{h})\}-\{y_j-\widehat{m}_{-j}(\bm{x}_j;\bm{h})\}}{b}\right),\label{eq:10}
\end{equation}
for $i=1,2,\dots,n$. It is likely that $y_j-\widehat{m}_{-j}(\bm{x}_j;\bm{h})=y_i-\widehat{m}_{-i}(\bm{x}_i;\bm{h})$, for some $j$, resulting in an unwanted term $\phi(0)/b$. A remedy is to exclude the $j$\textsuperscript{th} term that makes $y_j-\widehat{m}_{-j}(\bm{x}_j;\bm{h})=y_i-\widehat{m}_{-i}(\bm{x}_i;\bm{h})$ from the summation in~\eqref{eq:10}. Therefore, the density of $y_i$ is approximated by
\begin{equation}
\widehat{f}\big(\{y_i-\widehat{m}_{-i}(\bm{x}_i;\bm{h})\};b\big)=\frac{1}{n-n_i}\sum^n_{\substack{j\in J_i}}\frac{1}{b}\phi\left(\frac{\{y_i-\widehat{m}_{-i}(\bm{x}_i;\bm{h})\}-\{y_j-\widehat{m}_{-j}(\bm{x}_j;\bm{h})\}}{b}\right),\label{eq:22}
\end{equation}
where $J_i=\big\{j: y_j-\widehat{m}_{-j}(\bm{x}_j;\bm{h})\neq y_i-\widehat{m}_{-i}(\bm{x}_i;\bm{h}),\:\text{for}\: j=1,2,\dots,n\big\}$, and $n_i$ is the number of terms excluded from the summation in~\eqref{eq:10}. 

Given $\bm{h}^2$ and $b^2$, the likelihood of $\bm{y}=(y_1,y_2,\dots,y_n)^{\top}$ is the product of the density of $y_i$, for $i=1,2,\dots,n$, which are approximated by~\eqref{eq:22}. The likelihood is expressed explicitly as
\begin{equation*}
L\left(\bm{y}|\bm{h}^2,b^2\right)=\prod^n_{i=1}\left\{\frac{1}{n-n_i}\sum^n_{j\in J_i}\frac{1}{b}\phi\left(\frac{\{y_i-\widehat{m}_{-i}(\bm{x}_i;\bm{h})\}-\{y_j-\widehat{m}_{-j}(\bm{x}_j;\bm{h})\}}{b}\right)\right\}.
\end{equation*}

This likelihood function is not a proper likelihood since some terms are left out. Instead, the likelihood is a pseudo-likelihood. The pseudo-likelihood is a likelihood function associated with a family of probability distributions, which does not necessarily contain the true distribution \citep{GMT84}. As a consequence, the resulting Bayesian estimators, while consistent, may have an inaccurate posterior variance, and subsequent credible sets constructed by this posterior may not be asymptotically valid. Note that similar issues are discussed by, for example, \cite{CH03} and \cite{Mueller13}. There is no satisfactory solution to such a practical problem because on the one hand, dropping the terms $i=j$ invalidates posterior inference, and on the other hand, keeping all terms in likelihood would result in corner solutions for the bandwidths.

Nonetheless, the dropped terms may have a negligible impact on the resulting likelihood. If the dropped terms were included back to~\eqref{eq:22}, they would contribute to the would-be likelihood by $\phi(0)/b$, which is a constant and is unaffected by observations. This constant has no impact on the would-be likelihood when $\bm{h}$ is considered. Only when $b$ is considered, will the dropped terms have an impact on the would-be likelihood. However, this might be regarded as ``a cost" to avoid any possible corner solution for $b$. Any possible theoretical justification on this issue is left for future research.

\subsection{Priors}

As studied by \cite{Nadaraya64}, \cite{Watson64} and \cite{HM85}, for consistency of the nonparametric estimator, the bandwidth parameters $h_k$ and $b$ should go to infinity, as the sample size $n$ converges to infinity. In most cases, the optimal bandwidth rate can easily be derived up to a constant factor. We acknowledge that inverse Gamma prior density with constant hyperparameters may not always be appropriate for different sample sizes. As such, we choose prior densities by re-expressing bandwidths as the product of an unknown constant and a known order convergence rate:
\begin{align*}
& b=b_0\times n^{-1/5},\\
& h_k=h_{0,k}\times n^{-1/(d+4)}, \hspace{1cm} k=1,\ldots,d,
\end{align*}
where $b_0$ and $h_{0,k}$ do not depend on sample size $n$. Therefore, bandwidth estimation is equivalent to the estimation of $b_0$ and $h_{0,k}$. The priors of $b_0^2$ and $h_{0,k}^2$ can be chosen as the inverse Gamma prior with constant hyperparameters \citep[see, e.g.,][for discussion of other prior choices]{ZKH06, ZBK09}. Thus, the priors of $b_0^2$ and $h_{0, k}^2$ are given by
\begin{align*}
\pi\left(b_0^2\right)&=\frac{(\beta_b)^{\alpha_b}}{\Gamma(\alpha_b)}\left(\frac{1}{b_0^2}\right)^{\alpha_b+1}\exp\left\{-\frac{\beta_b}{b_0^2}\right\},\\
\pi\left(h_{0,k}^2\right)&= \frac{(\beta_{h})^{\alpha_{h}}}{\Gamma(\alpha_{h})}\left(\frac{1}{h_{0,k}^2}\right)^{\alpha_{h}+1}\exp\left\{-\frac{\beta_{h}}{h_{0,k}^2}\right\},
\end{align*}
respectively, where $(\alpha_b, \beta_b)$ and $(\alpha_{h}, \beta_{h})$ are hyperparameters. The hyperparameters were set at $\alpha_b=\alpha_h=1$ and $\beta_b=\beta_h=0.05$. These values are often used as the parameter values of an inverse Gamma density chosen as the prior of a variance parameter \citep[see, e.g.,][]{Geweke10}.

As pointed out by a reviewer, we may also consider bandwidths that are of the form $b=n^{-\omega}$ and $h=n^{-\eta}$, and choose to work with beta priors on the parameters $(\omega, \eta)\in (0,1)$. We assume that the priors of $\omega$ and $\eta$ are the beta density with hyperparameters denoted by Beta$(\psi_b, \kappa_b)$ and Beta$(\psi_{h_k}, \kappa_{h_k})$, respectively. Thus, the priors of $b$ and $h_k$ are given by
\begin{align*}
b &= n^{-\omega}, \qquad \pi_{\omega}(\psi_b, \kappa_b) = \frac{\omega^{\psi_b-1}(1-\omega)^{\kappa_b-1}}{B(\psi_b, \kappa_b)}, \qquad B(\psi_b, \kappa_b) =  \frac{\Gamma(\psi_b)\Gamma(\kappa_b)}{\Gamma(\psi_b+\kappa_b)}\\
h_k &= n^{-\eta_k}, \qquad \pi_{\eta_k}(\psi_{h_k}, \kappa_{h_k}) = \frac{\eta_k^{\psi_{h_k}-1}(1-\eta_k)^{\kappa_{h_k}-1}}{B(\psi_{h_k}, \kappa_{h_k})}, \qquad B(\psi_{h_k}, \kappa_{h_k}) = \frac{\Gamma(\psi_{h_k})\Gamma(\kappa_{h_k})}{\Gamma(\psi_{h_k}+\kappa_{h_k})},
\end{align*}
respectively, where $(\psi_b, \kappa_b)$ and $(\psi_{h_k}, \kappa_{h_k})$ are hyperparameters.

In simulation studies to be presented in Section~\ref{sec:3}, we choose our prior density to be the inverse Gamma prior density for $b_0$ and $h_{0,k}$. Because of the time-consuming nature of our simulation, we tend to leave the other prior choice for future research.

\subsection{Posterior}

According to Bayes' theorem, the posterior of $b^2$ and $\bm{h}^2$ is approximated by (up to a normalizing constant)
\begin{equation*}
\pi\left(b^2,\bm{h}^2|\bm{y}\right)\propto L\left(\bm{y}|b^2,\bm{h}^2\right)\pi\left(b^2\right)\prod^d_{k=1}\pi\left(h_k^2\right),
\end{equation*}
from which we use an adaptive random-walk Metropolis algorithm of \cite{GFS11} to sample $b^2$ and the elements of $\bm{h}^2$. As studied by \cite{GRG96}, \cite{RGG97}, \cite{RR01}, \cite{RR09} and \cite{LN18}, the acceptance rate in the adaptive random-walk Metropolis algorithm was controlled to be around 0.234 for multivariate parameters (i.e., bandwidths of predictors in the regression mean) and 0.440 for a univariate parameter (i.e., bandwidth in the error density). As Markov chain Monte Carlo iterations increase, the adaptive algorithm can automatically update the step size in a random-walk Metropolis algorithm, to achieve the optimal acceptance rates. If the proposal parameters are accepted in an iteration, then the step size is reduced in the next iteration to search the optimal parameters in greater detail. If the proposal parameters are rejected in an iteration, then the step size is increased in the next iteration to explore other possible values. The adaptive random-walk Metropolis algorithm is computationally fast with guaranteed acceptance rates, in comparison to a fixed step size in the classical Metropolis algorithm. 

When the sampling procedure is completed, the ergodic average of each simulated chain is used as an estimate of each element of $\bm{h}^2$ and $b^2$. Once bandwidths are estimated, the kernel-form error density can be derived according to~\eqref{errordensity}.

\section{Monte Carlo simulation studies}\label{sec:3}

The purpose of this simulation study is to examine the performance of the proposed Bayesian bandwidth estimation method by assessing the accuracy of the resulting local linear estimator and kernel density estimator. For comparison purposes, we also choose bandwidths through the ROT and CV methods described in \cite{Hardle90} and \cite{Scott92}. With bandwidths for the local linear estimator estimated through these three methods, we calculate the residuals, based on which we apply the ROT and likelihood CV methods to obtain the bandwidth for the kernel density estimator of errors. This is a two-step procedure, whereas the proposed Bayesian algorithm selects the bandwidths of the local linear estimator and kernel-form error density simultaneously.

\subsection{Simulation setup}

Consider the relationships between $y$ and $\bm{x}=(x_1, x_2, x_3)^{\top}$ given by
\begin{align}
  m_1(\bm{x})&=\cos(2\pi x_{1})+\sin(2\pi x_{1}), \label{eq:30} \\
  m_2(\bm{x})&=\sin(2\pi x_{1})+\cos(2\pi x_{2})+4(1-x_{3}^2).\label{eq:4}
\end{align}
A sample was generated by drawing $x_{1}, x_{2}, x_{3}$ independently from the uniform density on $(0,1)$. The errors were simulated from either $N(0, 0.5^2)$ or a mixture of two Gaussian densities defined as $0.7N(0,0.4^2)+0.3N(0,0.8^2)$. The response was calculated through \eqref{eq:30} and \eqref{eq:4} with the generated errors added. We consider three sample sizes: $n = 200, 1000$ and $5000$.

The relationships between the responses and explanatory variables were modeled by nonparametric regression models given (respectively) by
\begin{align}
y_j&=m(x_{j,1})+\varepsilon_j,  \quad j=1,2,\dots,n,\label{eq:5}\\
y_j&=m(x_{j,1},x_{j,2},x_{j,3})+\varepsilon_j, \label{eq:6}
\end{align}
where the errors are assumed to be iid with an unknown density.

\subsection{Simulation result}

The error density of~\eqref{eq:5} and~\eqref{eq:6} is assumed to be approximated by the kernel-form density given by~\eqref{errordensity}. With each pseudo sample generated from each model, we implemented the proposed Bayesian sampling algorithm to estimate bandwidths for local linear estimators of \eqref{eq:5} and~\eqref{eq:6}. The random-walk Metropolis algorithm was used to sample parameters, where the burn-in period contains the first 1000 draws, and the successive 10000 draws were recorded. 

The posterior averages of bandwidths for each model are presented in Table~\ref{tab:1}. The mixing performance of this posterior sampler is examined by the simulation inefficiency factor (SIF), which can be loosely interpreted as the number of draws needed to obtain independent draws from the simulated chain. For example, a SIF value of 20 indicates that, approximately, we should keep one draw for every 20 draws to derive independent draws \citep*[see, e.g.,][]{Roberts96, KSC98}.

\begin{table}[!htbp]
\centering
\renewcommand{\tabcolsep}{0.63cm}
\caption{Posterior estimates and statistics for Bayesian bandwidth estimation of one and three regressors under (mixture) Gaussian error density with $n=1000$ in each sample.}\label{tab:1}
\centering
\begin{tabular}{@{}lllclr@{}}\toprule
    Error   & Parameter & 95\% Bayesian     & Standard  & Batch-mean & SIF \\
    density & estimate  & credible interval & deviation & standard dev & \\\midrule
    $d = 1$ & \\\\
    Gaussian & $b\hspace{.09in}=0.2698$ & (0.1858, 0.3689) & 0.0474 & 0.0010 & 4.4 \\
             & $h_1=0.0607$ & (0.0514, 0.0706) & 0.0049 & 0.0001 & 4.5 \\
    \\
    Mixture & $b\hspace{.09in}=0.2958$ & (0.2196, 0.3835) & 0.0423 & 0.0009 & 4.3 \\
    Gaussian        & $h_1=0.0615$ & (0.0521, 0.0717) & 0.0050 & 0.0001 & 4.4 \\\midrule
    $d = 3$ & \\\\
    Gaussian & $b\hspace{.09in}=0.2653$ & (0.1790, 0.3660) & 0.0485 & 0.0011 & 5.4 \\
             & $h_1=0.0784$ & (0.0675, 0.0901) & 0.0058 & 0.0002 & 12.6 \\
             & $h_2=0.1608$ & (0.1308, 0.1930) & 0.0160 & 0.0006 & 15.0 \\
             & $h_3=0.0924$ & (0.0778, 0.1082) & 0.0078 & 0.0003 & 13.5 \\
    \\
    Mixture & $b\hspace{.09in}=0.2957$ & (0.2156, 0.3869) & 0.0442 & 0.0010 & 5.2 \\
    Gaussian        & $h_1=0.0802$ & (0.0688, 0.0923) & 0.0060 & 0.0002 & 12.3 \\
            & $h_2=0.1659$ & (0.1348, 0.1994) & 0.0166 & 0.0006 & 14.8 \\
            & $h_3=0.0951$ & (0.0800, 0.1114) & 0.0081 & 0.0003 & 13.3 \\\bottomrule
  \end{tabular}
\end{table}

The standard deviation of the posterior mean is approximated by the batch-mean standard deviation. It becomes smaller and smaller as the number of iterations increases and the sampler achieves a reasonable mixing performance. We used the SIF and batch-mean standard deviation to monitor the mixing performance. Table~\ref{tab:1} presents the values of these two indicators, which show that the sampler has mixed very well.

\subsection{Comparison of bandwidth estimation methods}

The accuracy of estimated bandwidth parameters is compared via the ISE of the resulting local linear estimator. The ISE is defined as
\begin{equation*}
\text{ISE}(\widehat{m})=\int^{a_1}_{a_0} \left[\widehat{m}(x;\bm{h})-m(x)\right]^2dx,
\end{equation*}
where $[a_0,a_1]$ is the support of $m(x)$. Given a set of grids, $a_0=x_1<x_2<\dots<x_{1000}=a_1$, we approximate the ISE by
\begin{equation*}
\widetilde{\text{ISE}}(\widehat{m}) = \frac{a_1-a_0}{1000}\sum^{1000}_{j=1}\left[\widehat{m}\left(x_j\right)-m\left(x_j\right)\right]^2.
\end{equation*}
The CV method aims to derive an optimal $\bm{h}$ that minimizes ISE \citep*[see, e.g.,][]{Hardle91,SBS94}. However, the difficulty in numerically deriving an optimal bandwidth increases with the dimension.

Another criterion for bandwidth selection is the mean ISE (MISE) given by
\begin{equation*}
\text{MISE}(\widehat{m})=\mathbb{E}\left\{\int^{a_1}_{a_0} \left[\widehat{m}(x;\bm{h})-m(x)\right]^2dx\right\},
\end{equation*}
where $\mathbb{E}\{\cdot\}$ denotes expectation. It is well known that the optimal bandwidth that minimizes MISE does not have a closed-form \citep{ZBK09}; in practice, it is usually approximated asymptotically. When data are observed from the multivariate normal density, and the kernel function is the standard Gaussian density, the optimal bandwidth that minimizes MISE can be approximated by
\begin{equation*}
  h_k=\sigma_k\left[\frac{4}{(d+2)n}\right]^{1/(d+4)},
\end{equation*}
where $\sigma_k=\min\{s_k,q_k/1.34\}$ with $s_k$ the sample standard deviation and $q_k$ the interquartile range for the $k$\textsuperscript{th} predictor, for $k=1,2,\dots,d$. This method is often used in practice because of its simplicity and a lack of alternatives \citep[see, e.g.,][]{Silverman86, Scott92, BA97}. However, we might be misled by the bandwidths selected through ROT when the true density is far from Gaussian distribution. The CV method given in~\eqref{eq:50} has been extensively discussed for bandwidth selection \citep[see, e.g.,][]{WW75, HM85, HM00}. In this simulation study, we consider the ROT and CV as competing methods for the Bayesian method.

\begin{figure}[!htbp]
\centering
\includegraphics[width = \textwidth]{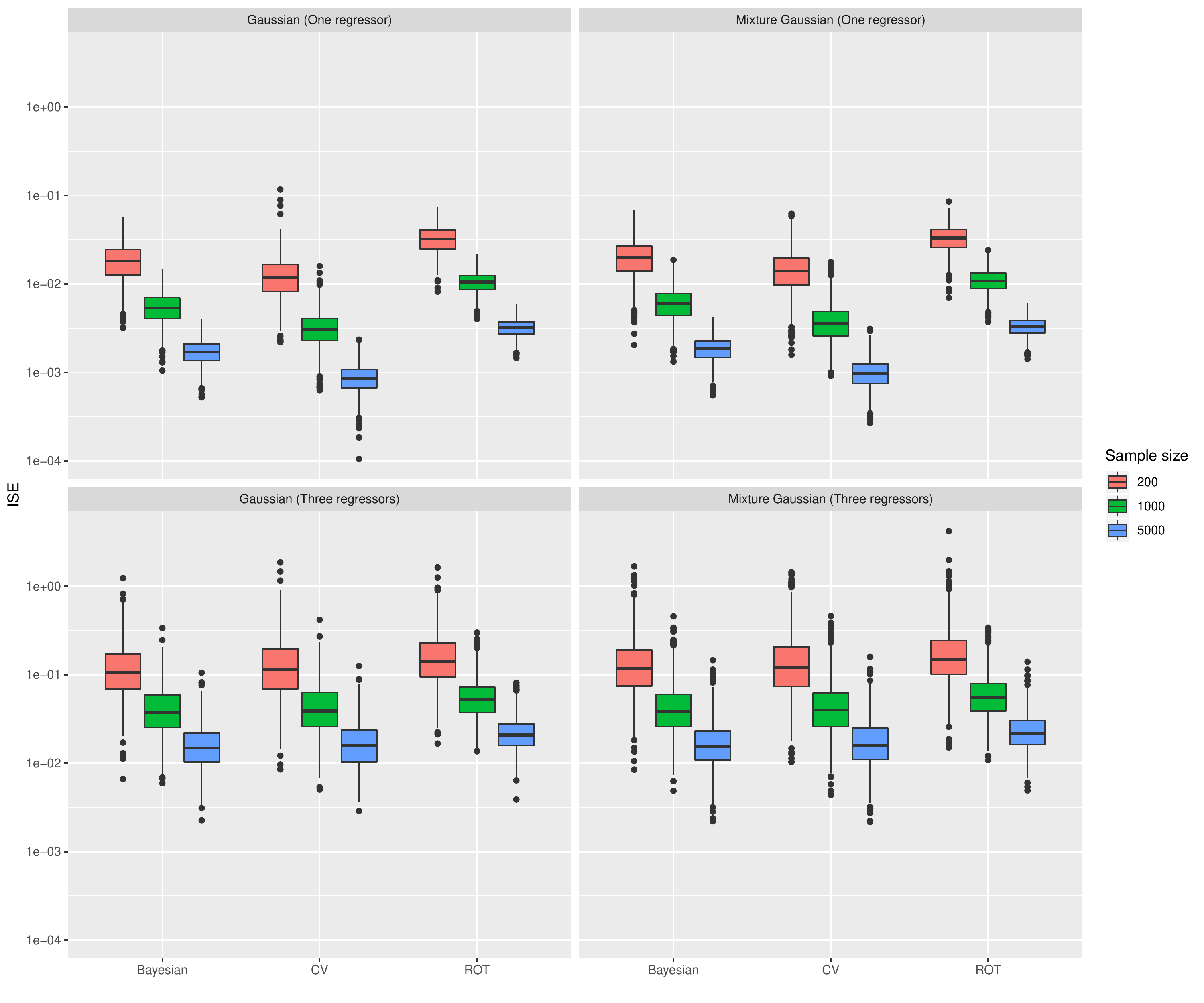}
\caption{Boxplots of 1000 ISE derived through each bandwidth estimation method over $n=200, 1000$ and $5000$ generated samples for estimating the regression function, where errors were simulated from the Gaussian density $\text{N}(0, 0.5^2)$ and mixture Gaussian density $0.7\text{N}(0,0.4^2)+0.3\text{N}(0,0.8^2)$ for both one and three regressors.}\label{fig:1}
\end{figure}

We calculate the ISE of the local linear estimator of the regression function with bandwidths estimated by the ROT, CV and Bayesian methods. The boxplots of ISE values are presented in Figure~\ref{fig:1}. When $d=1$, the CV has the best performance, with MISE performing worse than the other methods considered. However, as $d$ increases to three, the proposed Bayesian method performs the best.

\begin{figure}[!htbp]
\centering
\includegraphics[width = \textwidth]{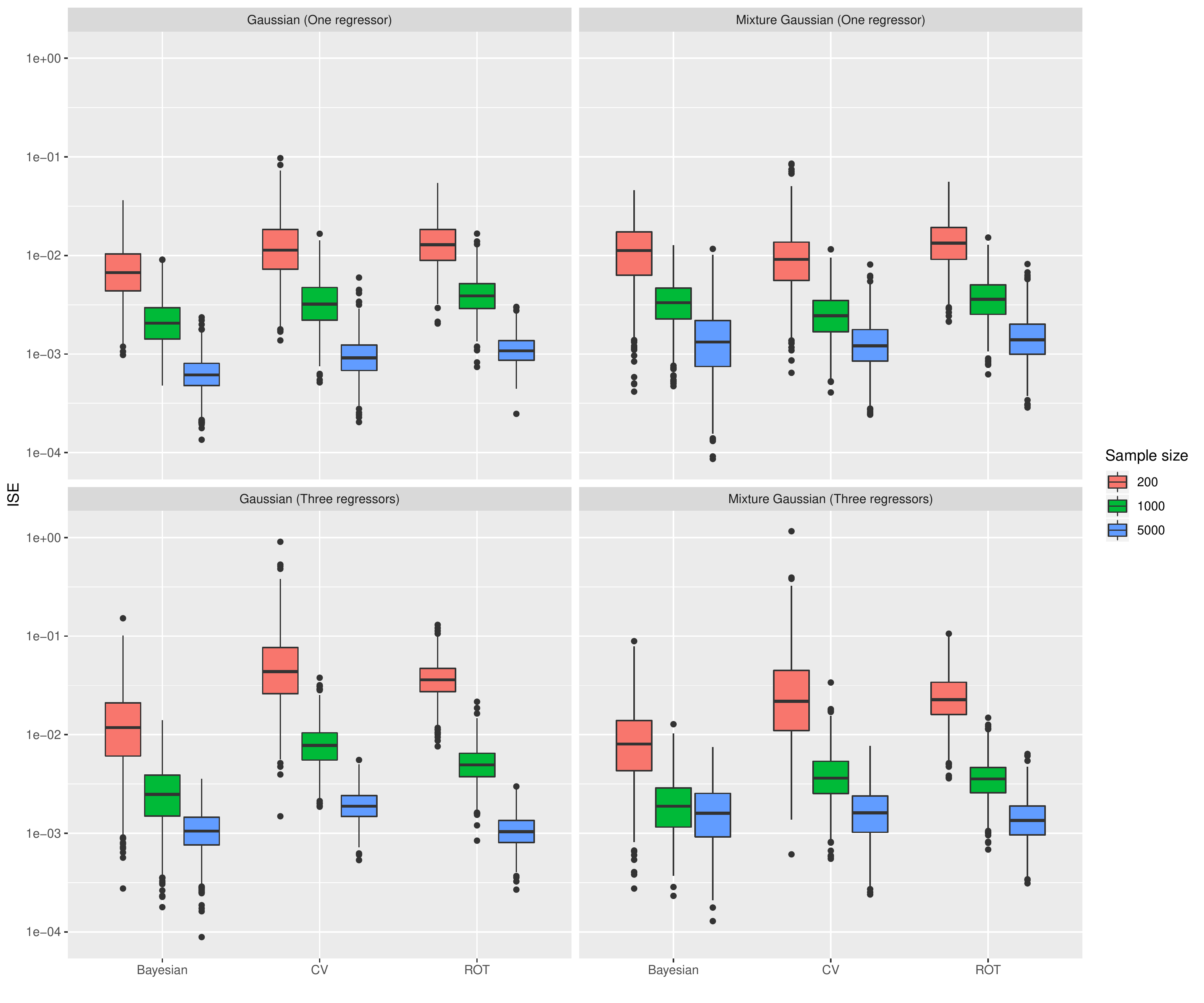}
\caption{Boxplots of 1000 ISE derived through each bandwidth estimation method over $n=200, 1000$ and $5000$ generated samples for estimating the error density, where errors were simulated from the Gaussian density $\text{N}(0, 0.5^2)$ and mixture Gaussian density $0.7\text{N}(0,0.4^2)+0.3\text{N}(0,0.8^2)$ for both one and three regressors.}\label{fig:2}
\end{figure}

After estimating the regression function, we apply the ROT and likelihood CV to obtain the bandwidth of the estimated error density via a two-step procedure in which the ROT and CV methods are applied twice separately. In contrast, our Bayesian algorithm simultaneously estimates bandwidths of regression function and kernel-form error density. Performance of the resulting bandwidth estimation for the kernel-form error density is summarized in Figure~\ref{fig:2}. When the errors are generated from the Gaussian density and mixture Gaussian density, the distribution of ISE values derived through the Bayesian method is generally smaller than those derived from either the ROT or CV method.

\subsection{Comparison between the local constant and local linear estimators}

\begin{figure}[!htbp]
\centering
\includegraphics[width = \textwidth]{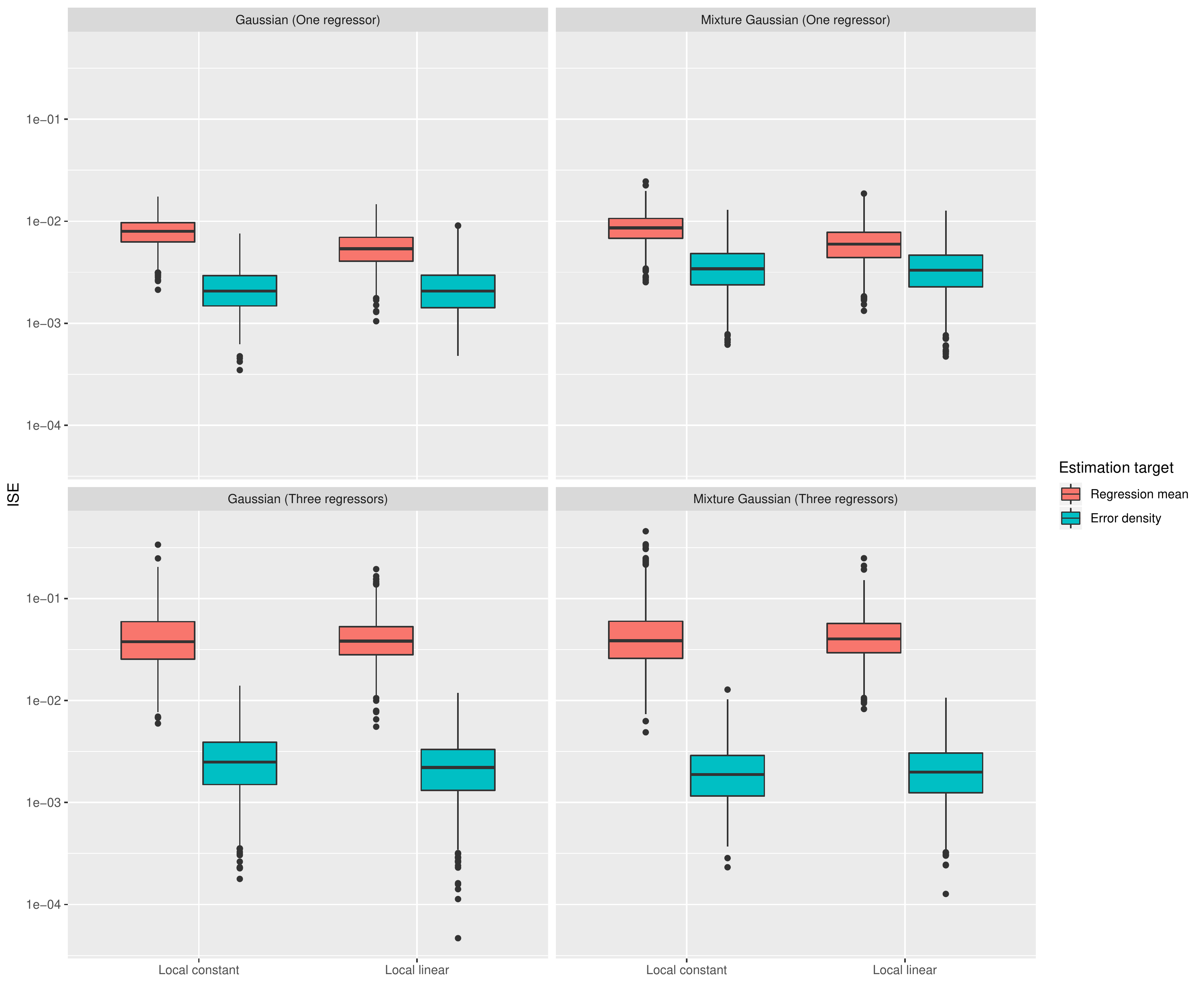}
\caption{Boxplots of 1000 ISE derived through the local constant and local linear estimators using the Bayesian bandwidth estimation method over $n=1000$ generated samples for estimating the regression function and error density, where errors were simulated from the Gaussian density $\text{N}(0, 0.5^2)$ and mixture Gaussian density $0.7\text{N}(0,0.4^2)+0.3\text{N}(0,0.8^2)$ for both one and three regressors.}\label{fig:3}
\end{figure}

We compare the proposed Bayesian bandwidth estimation method with the local linear estimator to the Bayesian bandwidth estimation with the local constant estimator introduced by \cite{ZKS11}. Although the two methods are both Bayesian, they differ in terms of the regression mean estimator. In Figure~\ref{fig:3}, we display boxplots of 1000 ISE values estimated by the local constant and local linear estimators. The local linear estimator improves the estimation accuracy of the regression mean function while maintaining similar values of ISE for the estimation of error density. 

Further, model selection in Bayesian inference is conducted through the marginal likelihood, which is the expectation of the likelihood with respect to the prior parameters. Here, we use the marginal likelihood to compare the two estimators and assess their statistical significance. The integral of the product of the likelihood and prior of parameters is rarely calculated analytically, but it is often computed numerically \citep[see, e.g.,][]{Chib95, Geweke99}.
  
Let $\theta = (\bm{h}, b)$ denote the parameter vector and $\bm{y}$ the data. The marginal likelihood reflects a summary of evidence provided by the data supporting one estimator as opposed to its competing estimator.  \cite{Chib95} shows that the marginal likelihood under an estimator $\cal A$ is expressed as
\begin{equation}
P_{\cal A}(\bm{y}) = \frac{l_{\cal A}(\bm{y}|\theta)\pi_{\cal A}(\theta)}{\pi_{\cal A}(\theta|\bm{y})}, \label{eq:marginal_likelihood}
\end{equation}
where $l_{\cal A}(\bm{y}|\theta), \pi_{\cal A}(\theta)$ and $\pi_{\cal A}(\theta|\bm{y})$ denote likelihood, prior and posterior under estimator $\cal A$. $P_{\cal A}(\bm{y})$ is computed at the posterior estimate of $\theta$. The numerator of~\eqref{eq:marginal_likelihood} has a closed form and can be computed analytically, the denominator of~\eqref{eq:marginal_likelihood} is the posterior of $\theta$, which is replaced by its kernel density estimator from the simulated chain of $\theta$ through a posterior sampler. The Bayes factor of estimator $\cal A$ against estimator $\cal B$ is defined as
\begin{equation*}
\text{BF} = \frac{P_{\cal A}(\bm{y})}{P_{\cal B}(\bm{y})} = \exp\{\ln P_{\cal A}(\bm{y}) - \ln P_{\cal B}(\bm{y})\},
\end{equation*} 
where $P_{\cal A}(\bm{y})>P_{\cal B}(\bm{y})$. The Bayes factor is used to decide whether estimator ${\cal A}$ is favored over estimator ${\cal B}$. According to \cite{KR95}, a Bayes factor value between 1 and 3 indicates that the evidence supporting estimator ${\cal A}$ against estimator ${\cal B}$ is insignificant. When the Bayes factor is between 3 and 20, estimator ${\cal A}$ is favored over ${\cal B}$ with positive evidence; when the Bayes factor is between 20 and 150, ${\cal A}$ is favored over ${\cal B}$ with strong evidence; and when the Bayes factor is above 150, ${\cal A}$ is favored over ${\cal B}$ with very strong evidence. In Table~\ref{tab:2}, based on 1000 repetitions, we compare the log marginal likelihood (LML) and Bayes factor between the local constant and local linear estimator and find a strong preference for the local linear estimator.  

\begin{table}[!htbp]
\centering
\renewcommand{\tabcolsep}{0.14cm}
\caption{Mean and standard deviation (sd) of 1000 LML derived through the Bayesian bandwidth estimation method with the local constant and local linear estimators averaged over 1000 generated samples for estimating the error density, where errors were simulated from the Gaussian density $\text{N}(0, 0.5^2)$ and mixture Gaussian density $0.7\text{N}(0,0.4^2)+0.3\text{N}(0,0.8^2)$ for both one and three regressors. The minimum values are highlighted in bold. Chib denotes the LML computed by \cite{Chib95}, while Geweke denotes the LML computed by \cite{Geweke99}.}\label{tab:2}
\centering
\begin{tabular}{@{}lrrrrrrrr@{}}\toprule
    & \multicolumn{4}{c}{$d=1$} & \multicolumn{4}{c}{$d=3$} \\
    & \multicolumn{2}{l}{Gaussian} & \multicolumn{2}{l}{Mixture Gaussian} & \multicolumn{2}{l}{Gaussian} & \multicolumn{2}{l}{Mixture Gaussian} \\\cmidrule{2-9}
Estimator        & Chib & Geweke & Chib & Geweke & Chib & Geweke & Chib & Geweke \\\midrule
LML (local constant) & -763.27 & -763.35 & -837.75 & -837.86 & -859.23 & -859.56 & -932.46 & -932.81 \\
LML (local linear) & -754.99 & -755.05 & -829.86 & -829.94 & -806.91 & -807.17 & -883.64 & -883.92 \\
Bayes factor & 3944 & 4024 & 2670 & 2752 & 5.28e+22 & 5.66e+22 & 1.59e+21 & 1.71e+21 \\
\bottomrule
\end{tabular}
\end{table}

\section{An application to firm ownership concentration}\label{sec:4}

\cite{YY03} present an empirical investigation on how expenditure on managerial activities inside a firm is affected by a firm's ownership concentration. Their data were collected from $n=185$ firms listed on the Japanese stock market. Using this data set, \cite{KKG11} investigate the relationship between a measure of expenses on managerial private benefits $(y)$ and three explanatory variables -- firm age $(x_1)$, ownership concentration $(x_2)$ and firm profit $(x_3)$ -- to justify their proposed varying-coefficient regression model. Note that the ownership concentration is represented by the percentage of shares possessed by the top ten shareholders. \cite{KKG11} find that all three explanatory variables have a significant impact on the response.

To illustrate our proposed method for bandwidth estimation, we considered the nonparametric regression model given by
\begin{equation*}
  y_i=m\left(x_{i,1},x_{i,2},x_{i,3}\right)+\varepsilon_i,\quad \text{for}\quad i=1,2,\dots,185,
\end{equation*}
where the error density is unknown and approximated by our proposed location-mixture density given by~\eqref{eq:22}.

To apply our proposed sampling algorithm, we assume that the priors of $b^2$ and $h_k^2$, for $k=1,2,3$, are the inverse gamma densities. As a sensitivity analysis, we also consider prior densities as the exponential densities, given by
\begin{equation*}
  \pi(z) = \tau \exp(-\tau z), \quad z\geq 0,
\end{equation*}
where the hyperparameter $\tau=1$. The estimated values of $b$, $h_k$ and their associated statistics are presented in Table~\ref{tab:3}. The batch-mean standard deviation and SIF values show that the sampler has reached good mixing performance.

\begin{table}[!htbp]
\renewcommand{\tabcolsep}{1cm}
\caption{Parameter estimates and statistics of Bayesian bandwidth estimation under mixture Gaussian error density for Japanese chemical industry data. The burn-in period contains the first 1000 draws, and the successive 10000 draws were recorded.}\label{tab:3}
\centering
\begin{tabular}{@{}llllr@{}}\toprule
     Bandwidth  & 95\% Bayesian & Standard & Batch-mean & SIF \\
     Estimate   &  credible interval & deviation & standard dev & \\\midrule
     \multicolumn{3}{l}{\hspace{-0.4in}{Inverse Gamma densities $(\alpha = 1, \beta = 0.05)$}} \\
     $b  \hspace{.086in} = 0.3026$ & (0.1998, 0.4304) & 0.0574 & 0.0010 & 5.7 \\
     $h_1 = 4.6976$ & (4.0505, 5.4587) & 0.3489 & 0.0077 & 9.7 \\
     $h_2 = 3.4029$ & (2.4674, 4.1056) & 0.4087 & 0.0098 & 11.6 \\
     $h_3 = 0.9467$ & (0.2345, 1.9318) & 0.4485 & 0.0175 & 30.3 \\
\\
     \multicolumn{3}{l}{\hspace{-0.4in}{Exponential prior densities $(\tau=1)$}} \\
     $b \hspace{.086in} = 0.3049$  & (0.1993, 0.4344) & 0.0585 & 0.0010 & 5.6 \\
     $h_1 = 4.6856$ & (3.9914, 5.3569)  & 0.3498 & 0.0079 & 10.1 \\
     $h_2 = 3.3958$ & (2.3863, 4.1494)  & 0.4195 & 0.0103 & 12.0 \\
     $h_3 = 0.9589$ & (0.2818, 1.9872)  & 0.4451 & 0.0173 & 30.0 \\\bottomrule
\end{tabular}
\end{table}

\subsection{In-sample fitting}

For comparison purposes, we compare the accuracy of bandwidth estimation for the local linear estimator obtained by the ROT and CV methods. By using the \textit{np} package \citep{HR08} in R \citep{Team11}, the resulting bandwidth vector of the local linear estimator is $(0.6288\times 10^8, 0.1315\times 10^{12}, 0.3184\times 10^{13})$, which is not a meaningful result. Consequently, the CV method fails to choose meaningful bandwidths. In contrast, the ROT and Bayesian methods both lead to reasonable bandwidth estimates, with bandwidths obtained by the ROT being $\bm{h}=(6.1720, 5.9416, 0.0233)^{\top}$.

\begin{figure}[!htbp]
  \centering
  \includegraphics[width=14cm]{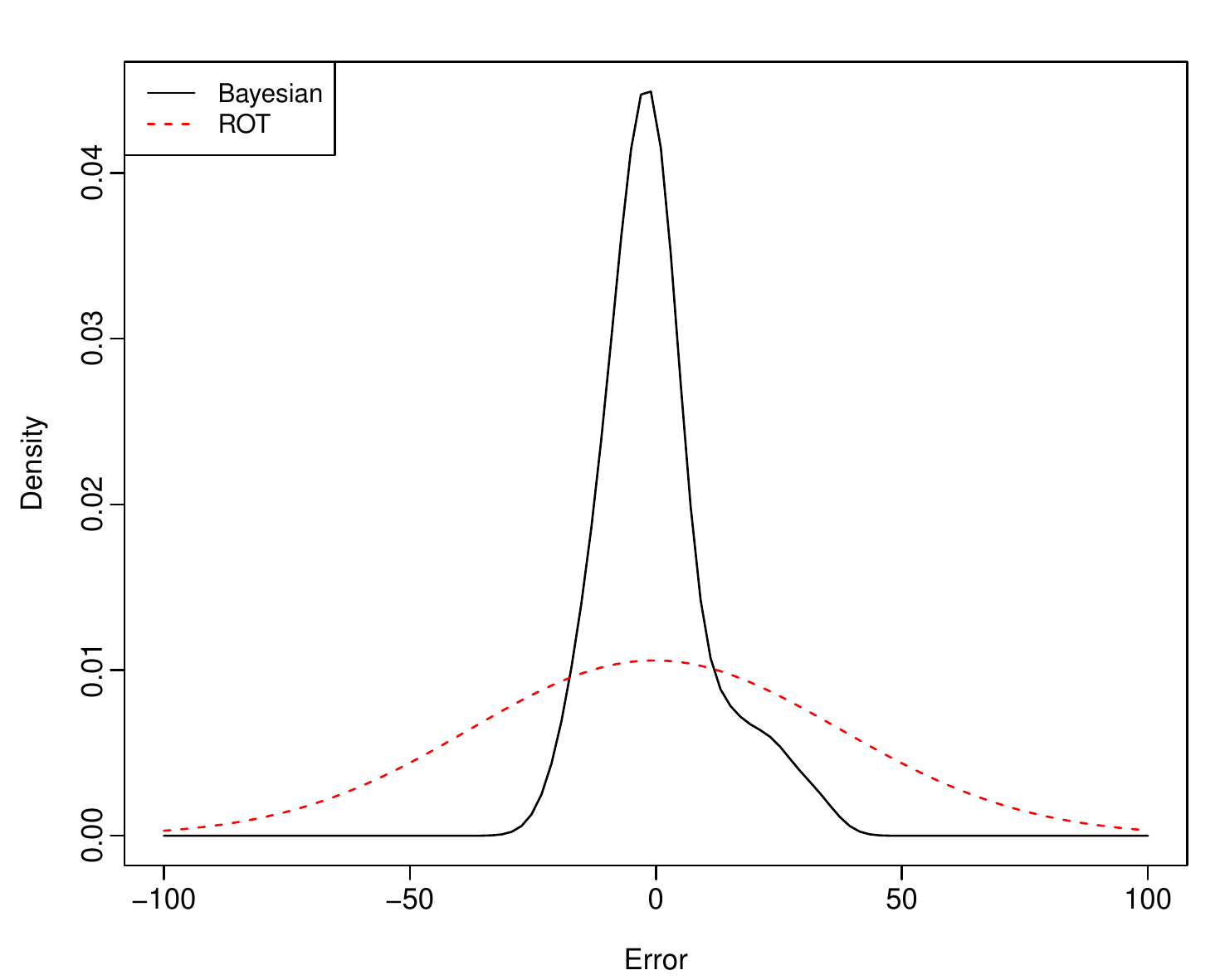}
  \caption{A plot of the estimated densities of regression errors.}\label{fig:4}
\end{figure}

Applying the ROT method to the kernel density estimator based on residuals, we obtained a bandwidth of $b=3.68$, which differs from the bandwidth derived through the proposed Bayesian method. The resulting difference in kernel density estimators can be seen in Figure~\ref{fig:4}.

\subsection{Out-of-sample prediction}

To assess the out-of-sample point forecast accuracy of the nonparametric regression estimator, we split the original sample into two sub-samples. The first one is called the learning sample, which contains the first 148 units $\{y_i, x_{i,1}, x_{i,2}, x_{i,3}\}_{i=1,\dots,148}$. The second one is called the testing sample, which contains the last 37 units $\{y_j, x_{j,1}, x_{j,2}, x_{j,3}\}_{j=149,\dots,185}$. The learning sample allows us to build the regression model with the estimated bandwidths, while the testing sample allows us to evaluate prediction accuracy.

To measure prediction accuracy, we consider the mean square forecast error (MSFE), mean absolute forecast error (MAFE) and mean absolute percentage error (MAPE). These are expressed as: 
\begin{align*}
\text{MSFE} = \frac{1}{37}\sum^{37}_{\omega=1}(y_{\omega} - \widehat{y}_{\omega})^2, \\ 
\text{MAFE} = \frac{1}{37}\sum^{37}_{\omega=1}|y_{\omega} - \widehat{y}_{\omega}|,\\ 
\text{MAPE} = \frac{1}{37}\sum^{37}_{\omega=1}\left|\frac{y_{\omega} - \widehat{y}_{\omega}}{y_{\omega}}\right| \times 100,
\end{align*}
where $y_{\omega}$ denotes holdout samples, while $\widehat{y}_{\omega}$ denotes the point forecasts. In Table~\ref{tab:4}, the MSFE and MAFE of the nonparametric regression estimator with bandwidths selected by the ROT and Bayesian bandwidth estimation methods are presented. 

\begin{table}[!htbp]
\centering
\tabcolsep 0.2in
\caption{Point forecast accuracy of the nonparametric regression with the local linear estimator, where the bandwidths are selected by the ROT and Bayesian methods.}\label{tab:4}
\begin{tabular}{@{}lccc@{}}
\toprule
Method  & MSFE & MAFE & MAPE \\
\midrule
ROT &  229.6831 & 11.7569 & 1.0836\%  \\
Bayesian &  197.8883 & 10.9798 & 1.0041\% \\
\bottomrule
\end{tabular}
\end{table}

\begin{figure}[!t]
\centering
\includegraphics[width=16cm]{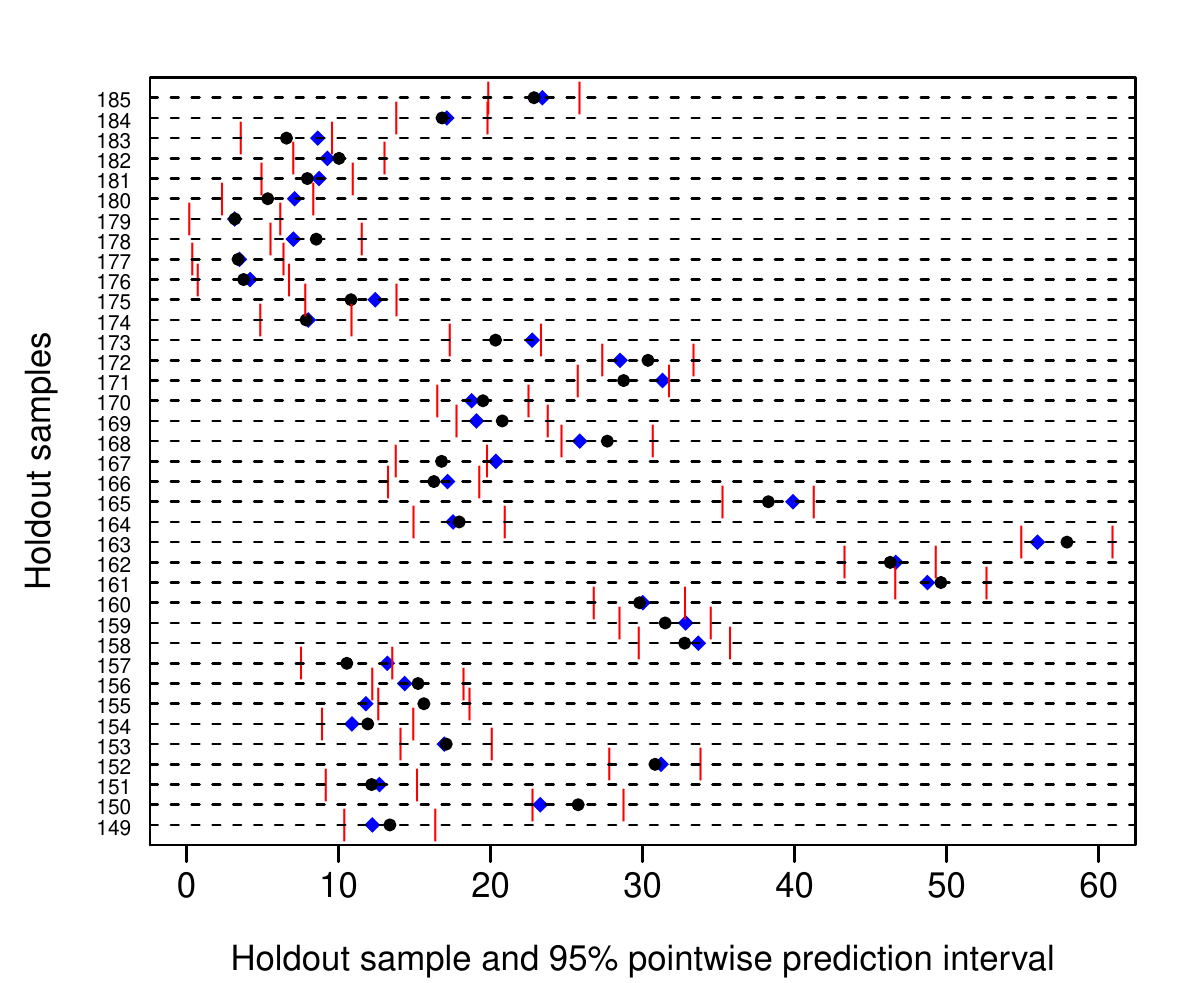}
\caption{A plot of holdout samples and the 95\% pointwise prediction intervals. The holdout data are shown as blue solid diamond-shaped dots. The point forecasts are shown as solid black circles, while the 95\% prediction intervals are shown as vertical bars. The empirical coverage probability is 94.59\%.}\label{fig:5}
\end{figure}

With the Bayesian approach, we can also compute the prediction interval nonparametrically. To this end, we first compute the cumulative density function (CDF) of the error distribution, over a set of grid points with a range of residuals. Then, we take the inverse of the CDF and find two grid points that are closest to the $\alpha/2$ and $(1-\alpha/2)$ quantile, where $\alpha = 0.05$ is the customarily level of significance. The $95\%$ pointwise prediction intervals are obtained by adding the two corresponding grid points to the point forecasts. For instance, the holdout samples of response are shown in solid diamond-shaped dots; the point forecasts are shown as solid black circles, while the $95\%$ pointwise prediction intervals are shown as vertical bars in Figure~\ref{fig:5}.

\section{An application to SPD estimation}\label{sec:5}

In the second application study, we consider a nonparametric regression model involving SPD estimation. \cite{AL98} show that in a dynamic equilibrium model, the price of a security is expressed as
\begin{equation*}
  P_t = \exp\left(\gamma_{t,\lambda}\lambda\right)\mathbb{E}_t^*[Z(S_T)]=\exp\left(\gamma_{t,\lambda}\lambda\right)\int^{\infty}_{-\infty}Z(S_T)f_t^*(S_T)dS_T,
\end{equation*}
where $T=t+\lambda$, $\lambda$ is the length of time to maturity, $\gamma_{t,\lambda}$ is a constant risk-free interest rate between $t$ and $T$, $\mathbb{E}^*_t$ is the expectation conditional on information available at date $t$, $S_T$ is the price of the security at date $T$, $Z(S_T)$ is the payoff of the security at the maturity date $T$, and $f_t^*(S_T)$ is the date $t$ SPD for the payoff of the security at date $T$.

When an option is of interest, the SPD is the second-order derivative of a call-option pricing formula with respect to the strike price calculated at $S_T$. According to \cite{AL98}, the date $t$ price of a call option is a nonlinear function of $\left(S_t, X_t, \lambda, \gamma_{t,\lambda}, \delta_{t,\lambda}\right)^{\top}$, which can be estimated through nonparametric regression, where $X_t$ is the strike price on a stock with date $t$ price $S_t$, and  $\delta_{t,\lambda}$ is the dividend yield.

To reduce the number of regressors, \cite{AL98} assume that the call-option pricing formula is given by the Black-Scholes (BS) formula, where the date $t$ volatility denoted by $\sigma_t$, is estimated by the nonparametric regression of the implied volatility on $\widetilde{z}_t=\left(F_t, X, \delta\right)$. Note that $F_t$ is the futures price of the underlying asset, $X$ is the strike price, and $\delta$ is the dividend yield. The kernel estimator of the regression function is
\begin{equation*}
  \widehat{\sigma}_t(F_t, X, \lambda|\bm{h}) = \frac{\sum^n_{j=1}K_{\bm{h}}\left(\widetilde{z}_t-\widetilde{z}_j\right)\widetilde{\sigma}_j}{\sum^n_{j=1}K_{\bm{h}}\left(\widetilde{z}_t-\widetilde{z}_j\right)},
\end{equation*}
where $\widetilde{\sigma}_j$ is the volatility implied by the price of the call option, and $\bm{h}=(h_1,h_2,h_3)^{\top}$ is a vector of bandwidths. According to \cite{AL98} and \cite{HKZ02}, the SPD is expressed as
\begin{align*}
  f_{\text{BS},t}(S_T)&=\frac{1}{S_T\sqrt{2\pi\widehat{\sigma}_t^2\lambda}}\exp\left\{-\frac{[\ln(S_T/S_t)-(\gamma_{t,\lambda}-\delta_{t,\lambda}-\widehat{\sigma}_t^2/2)\lambda]^2}{2\widehat{\sigma}_t^2\lambda}\right\}.
\end{align*}

The sample contains S\&P 500 options data from January 4 -- December 31, 1993, and the sample size is $n=14,431$. We apply the proposed Bayesian sampling algorithm to estimate bandwidths. Table~\ref{tab:5} presents the estimates of the bandwidths and their associated statistics. As a result, the bandwidth vector for the local linear estimator is $(0.9340, 4.7531, 9.4147)^{\top}$, which clearly differs from $(2.5336, 8.1416, 17.8335)^{\top}$ and $(0.2976, 7.0521, 3.4184)^{\top}$ derived through the ROT and CV methods. Moreover, the bandwidth for the kernel-form error density estimated through Bayesian sampling is 0.6134.

\begin{table}[!htbp]
\renewcommand{\tabcolsep}{0.94cm}
\caption{Estimates of the bandwidths and statistics: S\&P500 index options data. The burn-in period contains the first 100 draws, and the successive 500 draws were recorded.}\label{tab:5}
\centering
  \begin{tabular}{@{}lrccr@{}}\toprule
    Bandwidth   & 95\% Bayesian     & Standard  & Batch-mean     & SIF \\
    estimate    & credible interval & deviation & standard error &     \\\midrule
    $b \hspace{.086in} = 0.1132$ & (0.0981, 0.1350) & 0.0101 & 0.0014 & 9.5    \\
    $h_1 = 1.2887$   & (1.2592, 1.3067) & 0.0133 & 0.0040             & 44.5    \\
    $h_2 = 6.5582$   & (6.3092, 6.8044) & 0.1597 & 0.0525             & 54.1    \\
    $h_3 = 12.9901$  & (12.2074, 13.2636) & 0.3055 & 0.1003           & 53.8    \\\bottomrule
\end{tabular}
\end{table}

\begin{figure}[!htbp]
  \centering
  {\includegraphics[width=8.9cm]{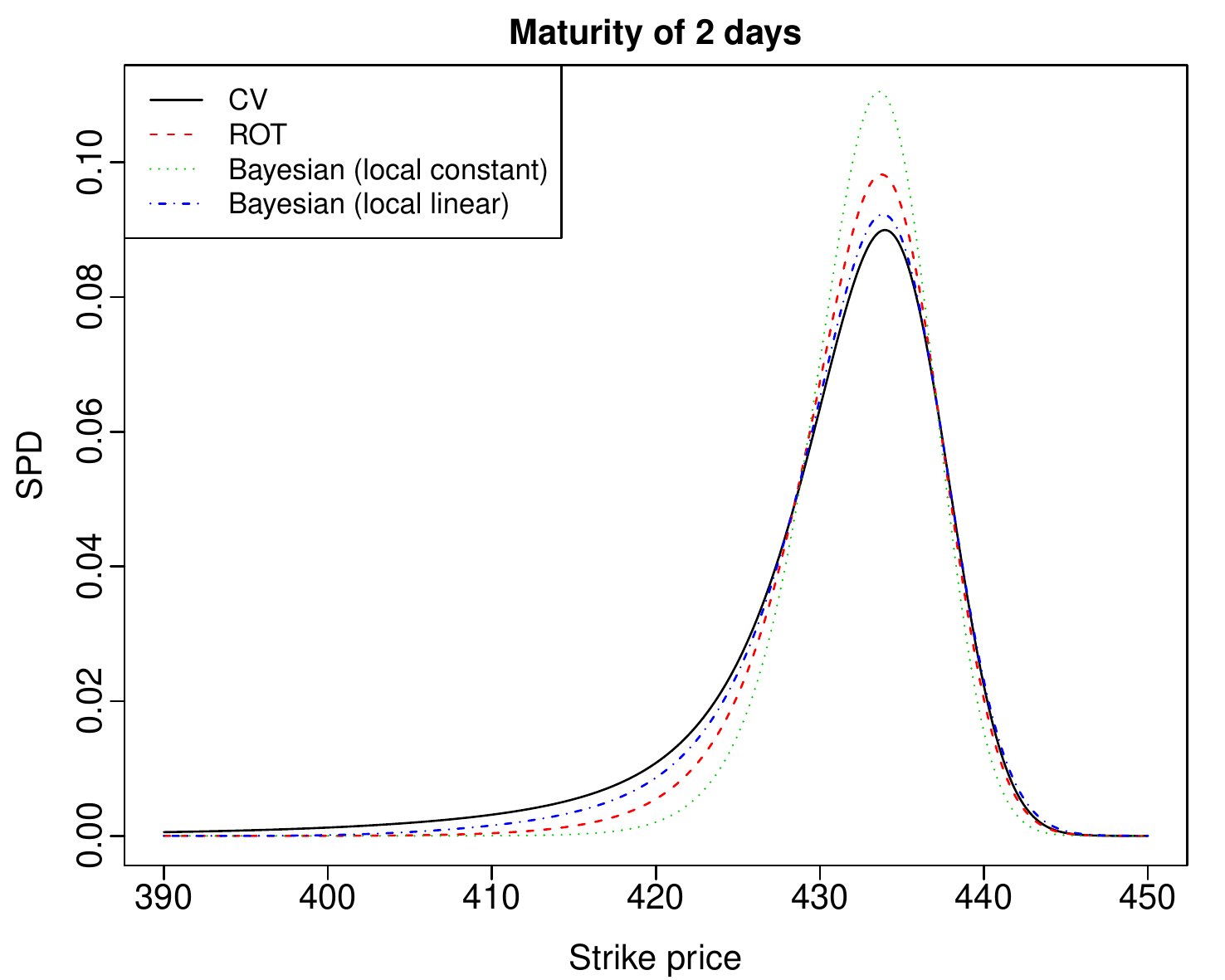}}
  \quad
  {\includegraphics[width=8.9cm]{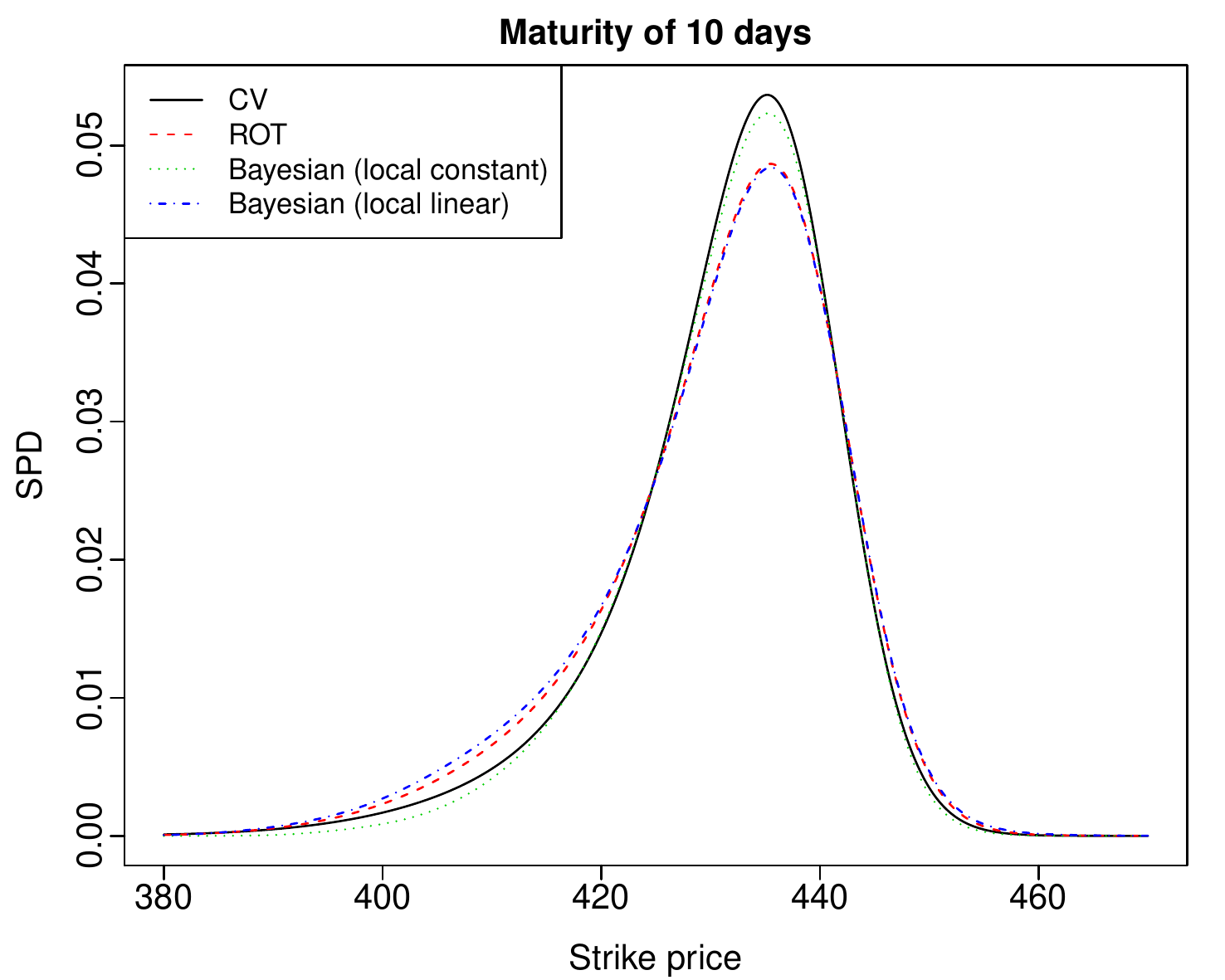}}
  \caption{Plots of the estimated SPD at maturities of 2 and 10 days based on S\&P 500 index options data.}\label{fig:6}
\end{figure}

Using the bandwidth vectors obtained from the ROT, CV and Bayesian methods, we plot the SPD graphs at maturities of two and ten days respectively, in Figure~\ref{fig:6}. At these maturities, the SPDs derived through Bayesian sampling are different from those derived through the ROT and CV methods, especially at the tail of the distribution. 

\section{Conclusion}\label{sec:6}

This paper presents a Bayesian sampling method to simultaneously estimate bandwidths for the local linear estimator of the regression function and kernel-form error density estimator in a nonparametric regression model, where the unknown error density is approximated by the mixture of Gaussian densities centered at individual residuals, and scaled by a bandwidth. We conducted a simulation study and demonstrated that the proposed Bayesian method outperforms ROT and CV in estimating bandwidths under the criterion of ISE. The proposed Bayesian method is able to provide not only bandwidth estimates but Bayesian credible intervals, which allow for assessing uncertainties associated with bandwidth estimates. Our Bayesian sampling procedure represents a data-driven solution to the problem of simultaneously estimating bandwidths for the kernel estimator of a regression function and kernel-form error density. It is well known that local linear fitting has many advantages over local constant fitting \citep[e.g.,][]{FG96}. As shown in Figure~\ref{fig:3}, this simple extension is necessary. 

For a nonparametric regression model of firm ownership concentration, we estimated bandwidths for local linear fitting; notably, the CV method failed to provide reasonable bandwidths. We also implemented the proposed Bayesian sampling algorithm to estimate bandwidths for the nonparametric regression model involved in SPD estimation. The SPD with bandwidths estimated through the proposed Bayesian sampling method differed from those with bandwidths via the ROT and CV methods. In comparison with ROT, our proposed method relaxes the assumption of a strict Gaussian error density to a mixture Gaussian error density. In comparison with CV, our proposed method utilizes prior information to constrain the range of optimal bandwidths, and consequently, a meaningful bandwidth estimate is obtained.

\vspace{.3in}
\bibliographystyle{agsm}
\bibliography{bandllf}

\end{document}